# Scientific mobility indicators in practice: International mobility profiles at the country level

Nicolas Robinson-Garcia, Cassidy R. Sugimoto, Dakota Murray, Alfredo Yegros-Yegros, Vincent Larivière and Rodrigo Costas

**Nicolas Robinson-Garcia** is a postdoctoral researcher at the *School of Public Policy* from the *Georgia Institute of Technology* in the United States. He holds a PhD in Social Sciences from the *University of Granada* (Spain). He is currently interested in the development of quantitative methods to analyze scientific mobility flows as well as the analysis of new data sources and methods to trace societal engagement of researchers.
*http://orcid.org/0000-0002-0585-7359*

*School of Public Policy*
*Georgia Institute of Technology, United States*
*elrobinster@gmail.com*

**Cassidy R. Sugimoto** is an associate professor of Informatics at the School of Informatics, Computing, and Engineering at Indiana University Bloomington. She conducts research within the domain of scholarly communication and scientometrics. She has published more than 70 journal articles on this topic. Her work has been presented at numerous conferences and has received research funding from the National Science Foundation and the Sloan Foundation, among other agencies. Dr. Sugimoto is actively involved in teaching and service and has thus been awarded with an Indiana University Trustees Teaching award (2014) and a national service award from the Association for Information Science and Technology (2009). She is currently President of the International Society for Scientometrics and Informetrics and a visiting professor at CWTS, Leiden University.
*https://orcid.org/0000-0001-8608-3203*

*School of Informatics and Computing*
*Indiana University Bloomington, USA*
*sugimoto@indiana.edu*

**Dakota Murray** is a PhD Student at Indiana University Bloomington studying Informatics under the Computing, Culture, and Society Track. His research interests aim at better understanding the structure of the global enterprise of science and advocate for positive change to create a more equitable scientific ecosystem. He is advised by and works as a research assistant for Dr. Cassidy Sugimoto. He has been funded by the EAGER grant sponsored by the NSF, and the IDEASc fellowship program sponsored by IMLS.
*https://orcid.org/0000-0002-7119-0169*

*School of Informatics and Computing*
*Indiana University Bloomington, USA*
*dakota.s.murray@gmail.com*

**Alfredo Yegros-Yegros** is a researcher at the Centre for Science and Technology Studies (CWTS) of Leiden University in the Netherlands. His current research revolves around quantitative studies of science and technology. More specifically, the analysis of public-private research interactions and knowledge flows, the study of science-technology linkages, and the study of methods potentially able to capture societal impact of scientific research are some of his research interests.

*Centre for Science and Technology Studies (CWTS)*
*Leiden University, The Netherlands*
*a.yegros@cwts.leidenuniv.nl*

**Vincent Larivière** is associate professor of information science at the École de bibliothéconomie et des sciences de l'information, l'Université de Montréal, where he teaches research methods and bibliometrics. He is also the scientific director of the Érudit journal platform, associate scientific director of the Observatoire des sciences et des technologies and a regular member of the Centre interuniversitaire de recherche sur la science et la technologie. He is also a visiting researcher at CWTS, Leiden University.
*https://orcid.org/0000-0002-2733-0689*

*École de bibliothéconomie et des sciences de l'information*
*Université de Montréal, Canada*
*vincent.lariviere@umontreal.ca*

**Rodrigo Costas** is a senior researcher at the Centre for Science and Technology Studies (CWTS) at Leiden University (the Netherlands). Rodrigo is also an Extraordinary Associate Professor at the Centre for Research on Evaluation, Science and Technology (CREST) of Stellenbosch University (South Africa). He holds a PhD in Library and Information Science from the CSIC in Spain. His areas of expertise include the fields of information science, scientometrics, and social media metrics. At CWTS he leads the research line in 'altmetrics', focused on developing new theoretical and analytical approaches to study the interactions between social media and science.
*http://orcid.org/0000-0002-7465-6462*

*Centre for Science and Technology Studies (CWTS)*
*Leiden University, The Netherlands,*
*and DST-NRF Centre of Excellence in Scientometrics and Science, Technology and Innovation Policy,*
*Stellenbosch University, South Africa*
*rcostas@cwts.leidenuniv.nl*

## Abstract

This paper presents and describes the methodological opportunities offered by bibliometric data to produce indicators of scientific mobility. Large bibliographic datasets of disambiguated authors and their affiliations allow for the possibility of tracking the affiliation changes of scientists. Using the Web of Science as data source, we analyze the distribution of types of mobile scientists for a selection of countries. We explore the possibility of creating profiles of international mobility at the country level, and discuss potential interpretations and caveats. Five countries—Canada, The Netherlands, South Africa, Spain, and the United States—are used as examples. These profiles enable us to characterize these countries in terms of their strongest links with other countries. This type of analysis reveals circulation among and between countries with strong policy implications.

Título en castellano: Indicadores de movilidad científica en acción: Perfiles de movilidad internacional a nivel de país

Resumen

Este trabajo presenta y describe las oportunidades metodológicas que ofrecen los datos bibliográficos para producir indicadores de movilidad científica. El uso de grandes conjuntos de datos bibliográficos con autores y afiliaciones desambiguadas, abren la posibilidad de rastrear cambios de afiliación de investigadores. Empleando la Web of Science como base de datos, desarrollamos distintas perspectivas para mostrar la movilidad observable de una selección de países. Exploramos la posibilidad de crear perfiles de movilidad internacional a nivel de países y discutimos cómo interpretar estos indicadores así como sus potenciales limitaciones. Para ello, estudiamos los casos de Canadá, Países Bajos, Sudáfrica, España y Estados Unidos. Sus perfiles no solo nos permiten identificar a grupos de investigadores que muestran distintos tipos de movilidad, pero también nos permiten caracterizar los países según aquellos otros con los que tienen mayores vínculos. Este tipo de análisis permiten realizar comparaciones entre países de origen y destino de cada uno de los países analizados, especialmente relevantes en el contexto de política científica.

**Palabras clave:** Movilidad científica, indicadores bibliométricos, movilidad internacional, internacionalización, política científica

## Introduction

Mobility of scientists is a topic of great concern in the science policy arena. In recent decades mobility has become a key issue due to an ever-more globalized research landscape. While some countries rely on foreign-born scientists to maintain their scientific status (**Levin; Stephen**, 1999), other countries envision mobility as a way to improve their national scientific capacities (**Ackers**, 2008), or to integrate themselves into a perceived elite of scientifically advanced countries (**Kato; Ando**, 2017). These examples align closely with the concept of *internationalization*, understood as 'the policies and practices undertaken by academic systems and institutions – and even individuals – to cope with the global academic environment' (**Altbach; Knight**, 2007). Therefore, many countries proactively implement policies to facilitate mobility of scientists (e.g., **Ackers**, 2005).

Until recently, bibliometrics has contributed very little to the study of scientific mobility. Existing scientific mobility indicators are often constructed using CV data, population statistics, or survey data (**Laudel**, 2003). These different data sources provide data at different levels of analysis; some sources dictate aggregate-level analysis (e.g., national census) while others provide for individual-level analysis (e.g., CV data). Examples of studies on mobility using such data sources are those by **Andújar; Cañibano; Fernández-Zubieta** (2015) using CV data, **Ackers** (2005), based on interviews, **Jonkers; Tijssen** (2008), combining CV and bibliometric data, or **Cruz-Castro; Sanz-Menéndez** (2010) who use survey data, among others. While insightful, retrieving data from these sources is time consuming, resource intensive, and provides a fragmented picture of scientists' mobility flows, rather than a global overview of the phenomenon.

It was **Laudel** (2003) who first suggested that affiliations found on scholarly papers could be employed as a more systematic way to track researchers' mobility by identifying institutional changes. At the time, such a task was still too time consuming (although maybe not as much as alternatives) as there was no direct link between the author's paper and their affiliation; hence, CV data and manual online searches were necessary to verify publications and identify information missing from bibliometric sources. Recent enhancement of bibliographic databases has eased this task significantly thanks to the introduction of author-affiliation linkages (from 2008 onwards in Web of Science). These new linkages are essential for the development of author name disambiguation algorithms, improving their capacity to identify the complete bibliographic output of individuals (**Caron; van Eck**, 2014; **Smalheiser; Torvik**, 2009).

The first attempt to track scientific mobility by bibliometric means was conducted by **Moed; Aisati; Plume** (2013). In this study and in their subsequent analyses (**Halevi; Moed; Bar-Ilan**, 2016; **Moed; Halevi**, 2014), the authors explored the use of Scopus' Author ID (the disambiguated set of authors from Scopus) to track institutional changes of scientists across countries. In these studies, they established the feasibility of using bibliographic data to track mobility, and compare international collaboration indicators with mobility, productivity and scientific impact. These and similar studies addressed the assumption that scientific international mobility is beneficial to scientific systems (**Wagner; Jonkers**, 2017) and analyzed the phenomenon from a *brain drain/brain gain* perspective. This perspective is often reflected in the terminology used to describe mobility. For instance, **Moed; Halevi** (2014) refer primarily to scientific *migration,* without considering other types of mobility. Similarly, **Robinson-Garcia** *et al.* (2016) discuss return rates of outgoing scholars without mentioning other types of mobility or considering benefits from potential scientific collaboration ties between sending and receiving countries due to scientific mobility. These studies do not envision mobility as a *brain circulation* phenomenon.

**Sugimoto; Robinson-Garcia; Costas** (2016) suggested that bibliometric analysis using a network perspective could more closely align with the brain circulation theory of the global scientific workforce. They linked countries with weights based on the number of shared scientific workers, defined as those scholars who had, at any given point in time, been affiliated to more than one country. **Sugimoto** *et al.* (2017) went beyond the network approach by establishing distinct patterns of scientific mobility, based on affiliation changes, and analyzing scientific impact between these patterns. They distinguished between 'migrants' and 'travelers' based on their affiliation path and the type of affiliation linkages they maintained (or did not maintain) with their 'country of origin', the country in which they first published. Based on these two mobility types, **Sugimoto** *et al.* (2017) determined that 72.7% of those bibliometrically-identified mobile scholars are travelers—people who are mobile but who never lose the affiliation country in which they published their first paper.

More recently, **Robinson-Garcia** *et al.* (2018) applied and expanded this taxonomy, developing four types of mobile scholars at the country level, distinguishing for each country those incoming migrants and travelers, along with outgoing migrants and travelers. In this study we delve into this typology, explaining how mobility indicators are built when based on bibliometric data and showcasing their potential use in research policy. We aim to describe in practical terms how this taxonomy can be useful when analyzing scientific mobility for a set of selected countries. As a proof-of-concept, we have purposively selected five countries with different types of migration profiles: Canada, The Netherlands, South Africa, Spain and the United States. This is considered an exploratory analysis to show the methodological and analytical possibilities of international mobility profiles of countries. The rest of the paper is structured as follows. First, we briefly describe our dataset. Then we explain how mobility indicators are constructed based on bibliographic data, pointing to caveats to interpretation and presenting our taxonomy of mobility indicators. Next, we develop the international mobility profile of five countries, namely: Canada, South Africa, Spain, The Netherlands, and United States. The aim of these profiles is not only to describe their scientific workforce in terms of international mobility, but also to identify with which countries they show the strongest ties.

## Data

Our approach uses the CWTS in-house version of the Web of Science, which includes all publications indexed in the database since 1980. We identify individuals by clustering publications based on the author name disambiguation algorithm developed by **Caron; van Eck** (2014). This algorithm uses a rule-based scoring system that links together documents that are likely authored by the same person by comparing bibliographic metadata of publications at four levels of analysis; author, article, publication and citation. When in doubt of a match, the algorithm is conservative and splits authors.

Based on this name disambiguation algorithm we define mobility events as changes in the country of affiliations within the output of each individual. As affiliation is one of the weighted fields, we expect the mobility changes found in our analyses to underrepresent total mobility. This approach may also fail to capture short-term mobility, such as temporary stays or research visits, when these events are not often recorded as affiliations on a resulting publication. For each individual and for a given year, we identified their affiliation type, differentiating between co-affiliated scholars (affiliated to more than one country in a single publication) and single affiliations (affiliated with only one country in a single publication). Furthermore, we distinguished between their country of origin and receiving countries. We assume that an individual's country of origin is that country in which they published their first paper (**Robinson-Garcia; Cañibano; Woolley; Costas**, 2016) while their receiving countries are any country with which the scholar held a subsequent affiliation (**Sugimoto** *et al.*, 2017).

## Building scientific mobility indicators based on bibliographic data

Full details of our taxonomy of mobility types are provided elsewhere (**Sugimoto** *et al.*, 2017; **Robinson-Garcia** *et al.*, 2018). In this section we will offer a basic explanation on how we track mobility through publications and define the different mobility types we developed. The first thing to note in this regard is that publication data offers only a *proxy* of mobility and does not necessarily track physical movements. Hence, scholars can be linked to one or more countries at the same time, which results in a multiplicative effect. Another caveat of this approach is that our capability to reliably identify mobile scholars is dependent on the visible productivity of these scholars in terms of their publications. This requirement has two direct consequences: on the one hand, we underrepresent mobility of scholars with small numbers of publications; on the other hand, publication delays and years with no publications represent holes in our data that may hinder trend analyses.

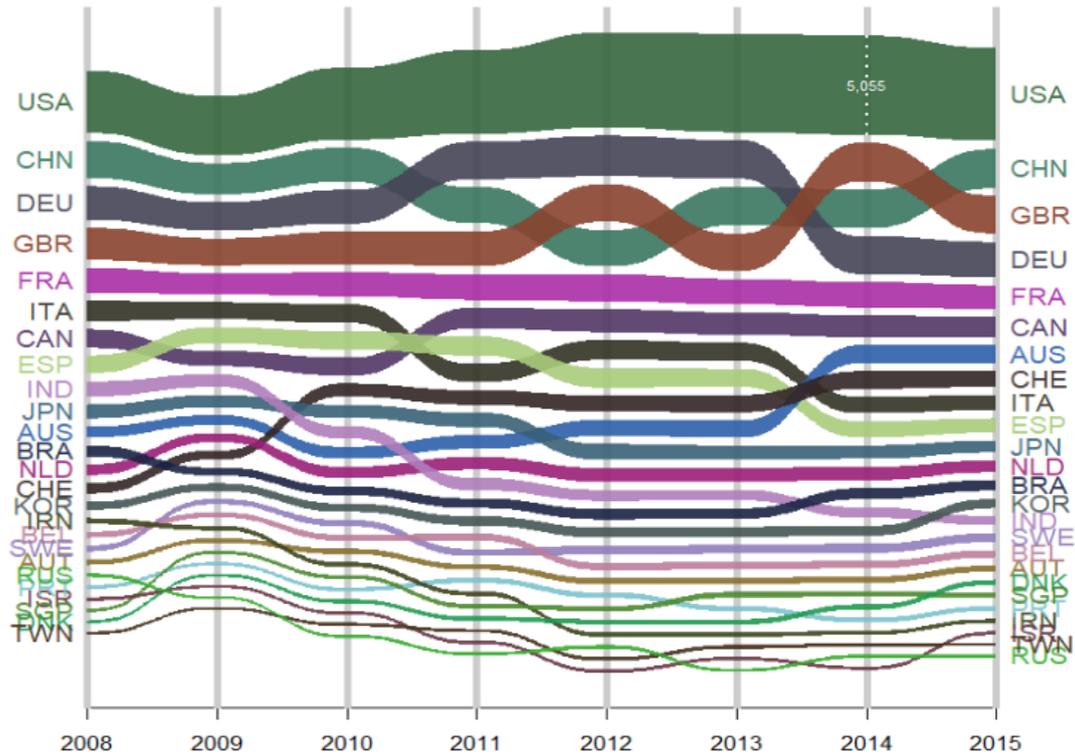

Figure 1. Trend analysis of number of affiliations for scholars publishing in Web of Science for the top 25 countries with the largest share of mobile scholars. Only scholars whose first publication year is 2008 and with at least 8 publications within the 2008-2015 period are included. ISO 3166-1 alpha-3 codes are used to name countries.

Figure 1 shows the number of affiliations by year for the top 25 countries with the largest share of mobile scholars who published the first paper in 2008 and who have published at least 8 papers within the 2008-2015 period. One of the problems with this type of analyses is dealing with blank years (years where a given scholars has produced no publications). One way of overcoming this limitation is to assume that no change takes place, and to assign to those empty years the country to which the scholar was last affiliated. Despite such issues, this approach provides insights, showing the dynamics of the scientific workforce of countries over time. For instance, we observe a decreasing number of mobile scientists in Italy, Spain, India or Russia. Similarly, we observe increases of internationally mobile scientific workforce in the United States, Australia or Switzerland.

Changes in the number of affiliations by country may result from scientists leaving one country and going to another, or because of scientists co-affiliating with a second or third country as well as their country of origin. Hence, some countries might see increases in the number of affiliations or the size of their workforce by recruiting scholars from elsewhere, while other countries may simply be *sharing* scholars with others. These two perspectives suggest that the phenomenon of scientific mobility is conceptually more closely related with a *brain circulation* framework (**Sugimoto; Robinson-Garcia; Costas**, 2017) rather than with the more reductionist *brain drain/brain gain* model. A circulation model obliges us to distinguish between different types of mobile scholars. Here we use the taxonomy presented by **Robinson-Garcia** *et al.* (2018) which defines four mobility types; at the country level, this taxonomy can be expanded to six types. For a complete description of the taxonomy we refer to the original paper. Figure 2 overviews the main differences between the four basic mobility types which we define as:

- **Not mobile.** These are scholars who have affiliation/s in a single country in all their publications. They represent 96.3% of the scholars in the 2008-2015 period (**Robinson-Garcia** *et al.*, 2018). This large share of apparent non-mobile scholars can be explained by the strong skewness in scientific productivity, with extremely large shares of scholars only authoring a few papers (**Ruiz-Castillo; Costas**, 2014).
- **Migrants.** These are scholars who, for at least one year, did not list the affiliation corresponding to their country of origin. They represent 1.0% of all scholars and 27.3% of mobile scholars in the 2008-2015 period (**Robinson-Garcia** *et al.*, 2018). At the country level, these scholars can be further characterized as:
    - **Emigrants.** Defined for country A as those who have country A as their country of origin and at any given time they cease being affiliated to it.
    - **Immigrants.** Defined for country A as those who have country B as their country of origin and at any given time are affiliated to country A.
- **Travelers.** Defined as scholars who at some point are affiliated to more than one country, but who in all years produce at least one publication still affiliated to their country of origin. They represent 1.3% of all scholars and 35.9% of mobile scholars in the 2008-2015 period. At the country level, these scholars can be further characterized as:
    - **Outgoing travelers.** Defined for country A as those who have country A as their country of origin and at any time they are also affiliated to country B while retaining their affiliation to country A.
    - **Incoming travelers.** Defined for country A as those who have country B as their country of origin and at any time they are also affiliated to country A while retaining their affiliation to country B.
- **Non-directionals.** Scholars who are affiliated in their first publication year to more than one country (they have more than one country of origin) and always show linkages between the same countries, hence precluding identifying directionality of changes. They represent 1.4% of all scholars and 36.8% of mobile scholars in the 2008-2015 period.

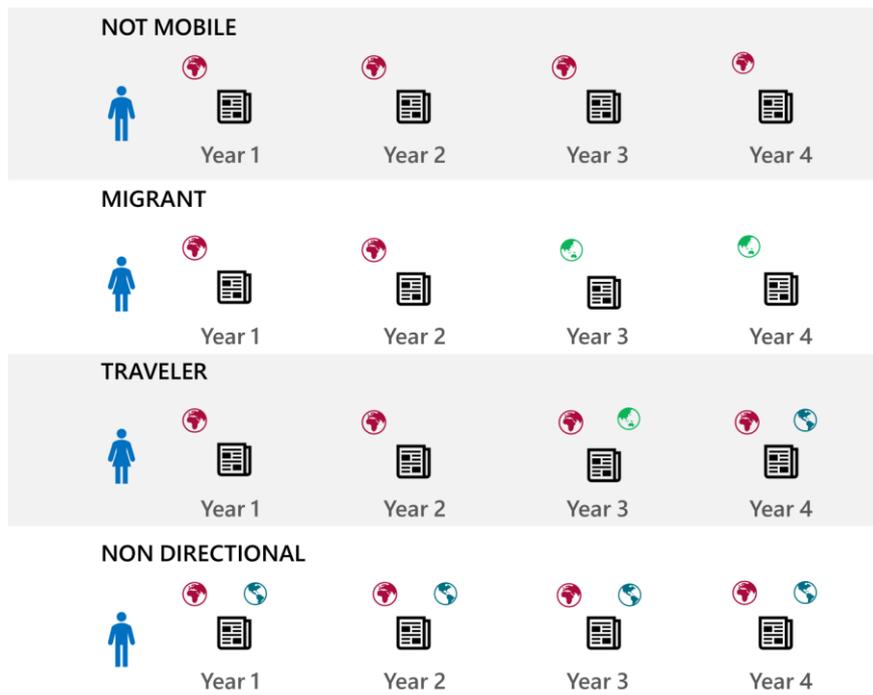

Figure 2. Overview of mobility types based on their affiliation changes. Red (Europe/Africa), green (Asia/Oceania), blue (America), to indicate variations in countries.

## The international mobility profiles of selected countries

This mobility taxonomy allows us to develop indicators at the country level, such as a country's share of mobile scientists, migrants, or travelers (**Robinson-Garcia** et al., 2018). In addition, citation impact or collaboration indicators can be calculated and compared for each distinct mobility population. Moreover, countries can be characterized and profiled based on their mobile scientific workforce.

Table 1 presents an overview of the internationally mobile scientific workforce of the five selected countries. As expected, the United States shows the largest overall number of scientists. Following the U.S. are Canada, Spain, The Netherlands and, lastly, South Africa. The United States shows the lowest share of mobile scientists (6.8%) while The Netherlands (16.7%) has the largest share of mobile scholars, followed by South Africa (15.0%). Considering only mobile scholars, the non-directionals are the most common type, accounting for, at the high end, 36.5% of all mobile scholars in the United States to a low of 25.5% in Spain. This group of scholars is the most difficult to explain as it is mostly formed by scientists with a single publication in which they are co-affiliated to more than one country. As suggested elsewhere (**Robinson-Garcia** et al., 2018), a portion of those identified as non-directionals may also be publications for which the algorithm did not cluster with the actual author. The second largest type of mobile scientists are travelers, ranging from 43.2% of all mobile scholars in Spain to 35.9% in the United States. Migrants are the least common type of mobile scholar, ranging between 34.1% of mobile scholars in Canada to 26.0% in South Africa.

We observe larger differences between these countries when comparing shares of incoming scholars (immigrants and incoming travelers) and outgoing scientists (emigrants and outgoing travelers). For this analysis, we omit non-directional scholars, as it is not possible to discern their directionality of movement. We find that 41.2% of mobile scientists in Spain are outgoing while 33.3% are incoming. At the other end of the spectrum we find South Africa, where 24.9% of their mobile scientists are outgoing while 43.8% are incoming scholars. In this regard, we observe that Spain is the only country with more outgoing scholars than incoming, while the rest show greater shares for incoming scholars.

Table 1. General overview of number of total number of scholars and by mobility type for Canada (CAN), Spain (ESP), Netherlands (NLD), United States (USA) and South Africa (ZAF) in the 2008-2015 period

|  | **CAN** | **ESP** | **NLD** | **USA** | **ZAF** |
| --- | --- | --- | --- | --- | --- |
| **Mobile scholars** | 54049 | 35418 | 30984 | 246388 | 8433 |
| **Emigrants** | 8743 | 6162 | 4635 | 31395 | 830 |
| **Immigrants** | 9668 | 4925 | 4656 | 36467 | 1366 |
| **Trained travelers** | 8375 | 8428 | 6117 | 37542 | 1268 |
| **Recruited travelers** | 11126 | 6863 | 6343 | 50979 | 2328 |
| **Non-directionals** | 16137 | 9040 | 9233 | 90005 | 2641 |
| **Total** | **430448** | **414999** | **185948** | **3641450** | **56360** |

Figure 3 shows the relation between mobility types and the directionality of scholars. Except for Canada, there is a larger share of outgoing travelers, those who acquire new links with other countries, than emigrants for each analyzed country. A similar trend is observed for incoming scholars, where for every country there is a greater share of incoming travelers than there are immigrants.

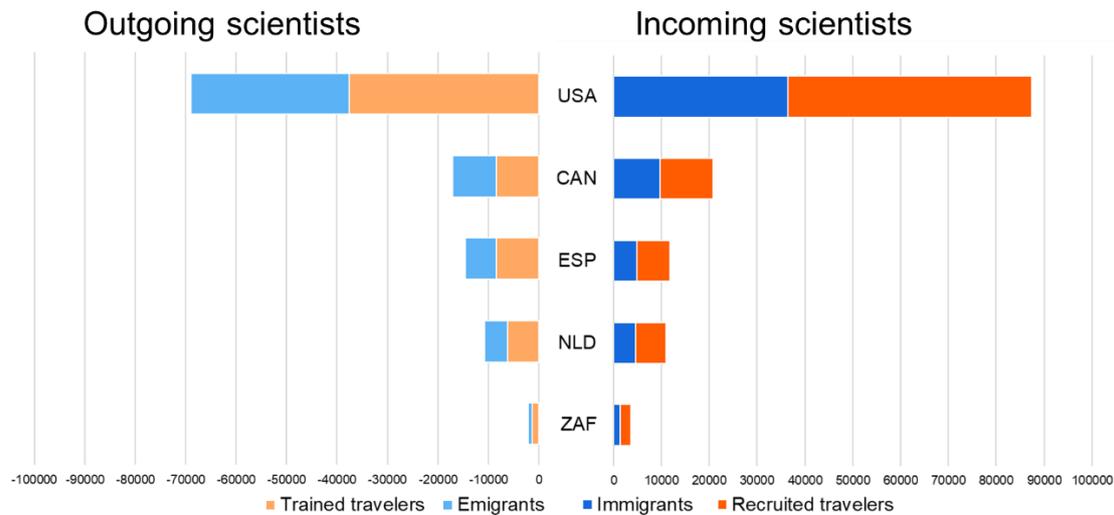

Figure 3. Total number of mobile scientists (excluding non-directionals) for Canada, Netherlands, South Africa, Spain and United States in the 2008-2015 period

In addition to analyzing the distribution of mobile scientists by country, we can also examine the source and destination countries of these mobile scholars. Figure 4 shows the flow of outgoing scholars originating in Spain in 2008, while figure 5 shows the flow of incoming scholars who eventually affiliate with Spain in 2015, starting from the top ten origin countries in 2008. These alluvial figures can be used to show the movements of scholars across countries and over time; still, there are many issues that must be considered when interpreting these figures. First, only a fixed set of scholars can be analyzed. Hence, we showcase only those scholars who started publishing in 2008 and ignore the population dynamics resulting from scholars entering and exiting the publishing system. Furthermore, these figures include all types of mobile scholars, which is problematic because, as discussed previously, co-affiliations have a multiplicative effect whereby one person can count towards multiple countries. Co-affiliation affects mostly (but not exclusively) travelers and non-directionals. To solve this issue, we use fractional counting of scholars by country and year—in other words, scholars are fractionalized by the countries with which they were affiliated during the relevant time period.

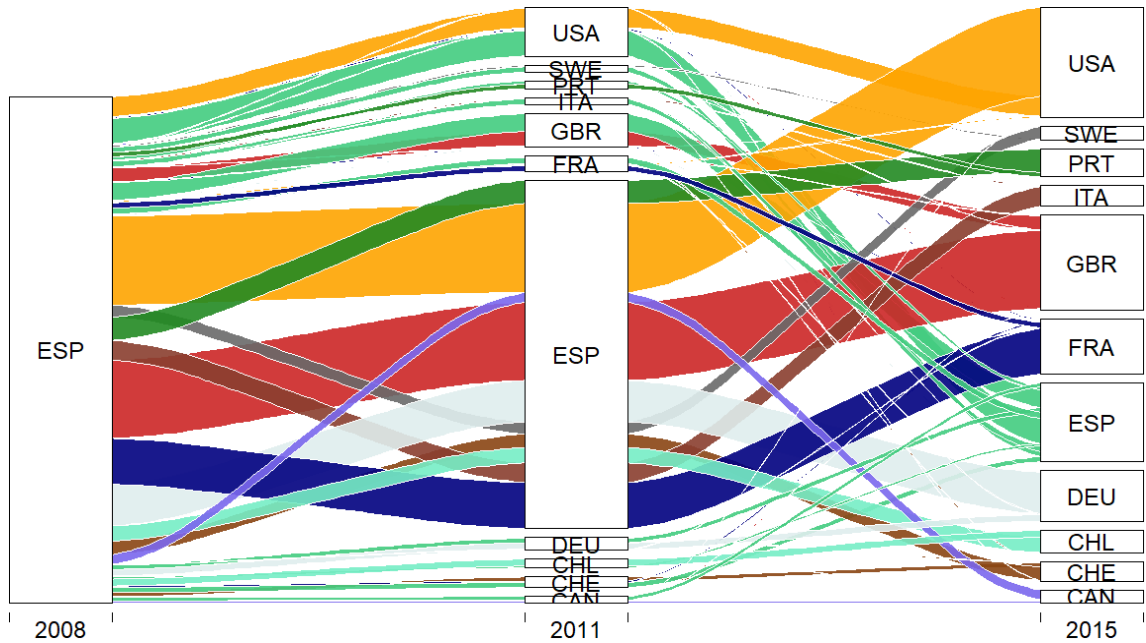

Figure 4. Flows of outgoing scholars from Spain. Only those who published their first paper in 2008 in Spain are included. Only the top 10 destinations are included.

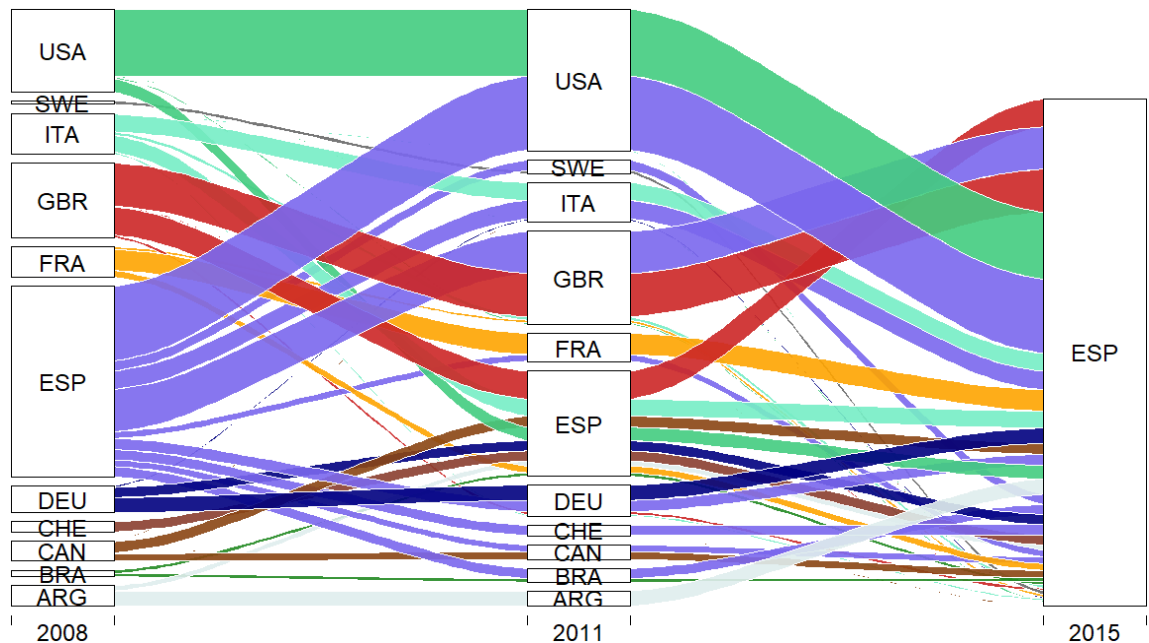

Figure 5. Flows of incoming scholars to Spain. Only those who published their first paper in 2008 and ended in 2015 in Spain are included. Only the top 10 countries of origin are included.

Most of the top 10 destinations of Spanish scholars are European countries (Figure 4) except for the United States, Chile and Canada. Using year of first publication as a proxy for age (**Nane; Larivière; Costas**, 2017), we can see that many outgoing scholars are young, and are thus likely to move to a different country after their third year of academic life, which may account for the lower share of Spanish affiliations in 2015. The top three destinations for scholars originating in Spanish are the United Kingdom, United States, and France. In the case of incoming scholars (Figure 5), we observe that the largest share are actually returned scholars, originating in Spain in 2008, becoming internationally mobile, and returning to Spain in 2015. Scholars from the United States and United Kingdom comprise the next largest shares of incoming scholars.

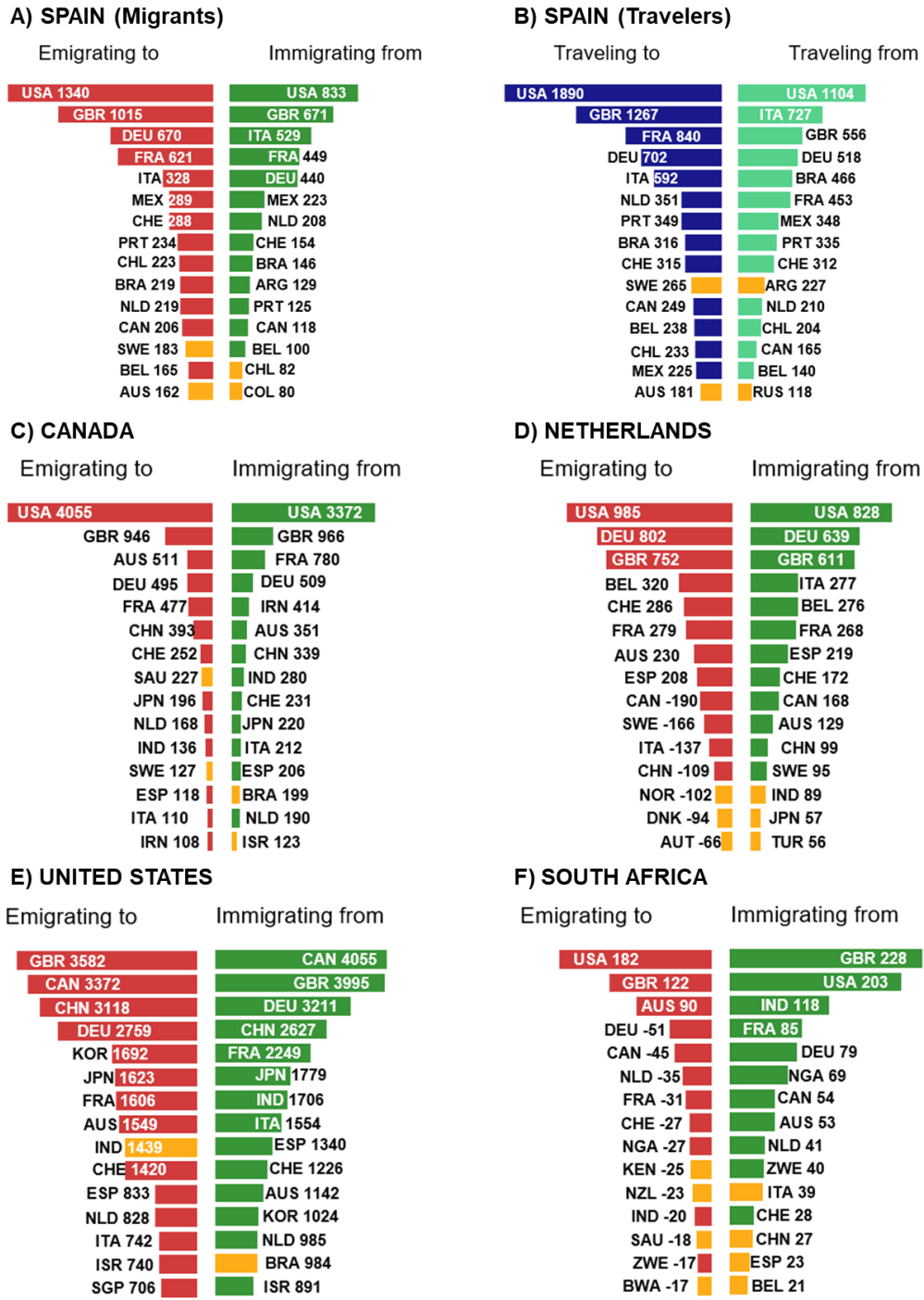

Figure 6. Scientific international mobility at the country level for A), B) Spain, C) Canada, D) Netherlands, E) United States and F) South Africa. Only for Spain migrants (A) and travelers (B) are shown, for the rest only migrants are included. Countries in yellow appear only in one of the top 15 list of linked countries to the analyzed one.

Whereas these alluvial visualizations give a quick overview of the flows and changes of the scholarly workforce over time, the analysis by mobility types provides a deeper understanding of a nation's mobility. Analysis of national mobility type distributions, when coupled with contextual information, can also reveal trends and directional flows which are hidden from the alluvial charts. Figure 6 shows the international mobility profile of the five analyzed countries. For Spain, we show the top countries with which most scholars are linked to, by mobility type. For the rest of the countries we show only the distributions for migrants. There are several countries that are both prominent sources of incoming and destinations for outgoing scholars and that are common between all five of our cases: these are the United Kingdom, the United States, Germany, and France. However, we do observe strong differences between each of the selected county's mobility distributions.

For instance, countries from the south of Europe such as Italy or Portugal, along with South American countries (i.e., Brazil, Mexico) are within the top countries sharing mobile scholars with Spain. However the outgoing distribution tends to include northern European countries such as Sweden or others like Australia as preferred destinations, rather than the southern European and South American countries common to incoming scholars.

In the case of Canada, we observe that United States is the preferred destination for most migrants, whereas the number of travelers is more distributed across other countries like the United Kingdom and Germany. We also observe the presence of Middle-East Asian countries like Iran and Saudi Arabia (only for emigrants), suggesting that these scholars could be receiving graduate training in Canada before returning to their home country. In the case of The Netherlands, there seems to be a more distributed share of scholars per country, although incoming scholars more often originate from southern European countries. The United States shows strong linkages with China, Germany, the United Kingdom, and with its neighbor Canada. Finally, in the case of South Africa, it is notable that while intra-continental migration is readily found in top positions, the United States and United Kingdom are the most preferred destinations and origins of migrants.

## Concluding remarks

This paper presents and describes a methodological approach to develop scientific mobility indicators based on bibliometric data. It delves into the possibility of using affiliation data from publications to track international scientific exchanges. We discuss the strengths and limitations of this approach and further describe a taxonomy of mobility types, which then can be used to create mobility profiles of countries. To this end, we have profiled the five countries to which the authors of this paper are or have been affiliated. We compare these profiles and observe prominent similarities and differences between each country's mobile scholarly workforce. Their profiles suggest that there is a selected group or 'elite' of countries, - namely United States, Germany, United Kingdom and France, - to which most of the selected countries are linked through mobility ties (Figure 6); this was already noted by **Sugimoto** *et al.* (2016; 2017).

It is important to highlight that the analytical approaches presented here are applicable to any country, and are possible to apply using any bibliographic database in which author-affiliation linkages are available and complete. Moreover, these methods can also be applied to any set of scholars; one example is analysis of a selection of countries, as done in this paper, but this same approach can also be used for any selection of regions, cities, or even institutions. It would even be possible to study the institutional mobility profile of the scholars affiliated with a given set of universities. Thus, the type of analysis that we introduce here is not restricted to countries, but is applicable to many different geographical and institutional entities. Future research will focus on these other more advanced and more granular scientific mobility profiles.

Scientific mobility indicators opens the door for analysis of global mobility trends and study of the evolution of the global scholarly workforce. At the same time, these indicators can provide a better understanding of the phenomenon of international collaboration (**Chinchilla-Rodriguez et al.**, 2017). Moreover, because they are built on bibliometric data, mobility indicators can easily be combined with citation impact indicators (**Sugimoto** et al., 2017), allowing the possibility for further developments and a more nuanced understanding of mobility. However, these indicators are not free of caveats and limitations, which must be considered both, when constructing and interpreting them. Our distinction between migrants and travelers can contribute to the ongoing discussion on mobility, as it reflects the complexity of the phenomenon. The distinction also goes beyond the common perception of scholarly mobility in science as a physical act or a permanent move. The fact that scholars may be contributing to more than one institution/country with their publications reveals that the current research context allows them to establish ties with different countries beyond physical mobility. Further research should focus on expanding the theoretical interpretation of such indicators to provide more advanced research policy discussions around mobility.

**Wagner, Caroline S.; Jonkers, Koen** (2017). Open countries have strong science. *Nature*, v. 550, n. 7674, pp. 32-33
*http://doi.org/10.1038/550032a*